\documentclass[a4paper]{article}

\usepackage{graphicx}
\usepackage{hyperref}
\usepackage{subfig}
\usepackage{csquotes}
\usepackage{INTERSPEECH2022}

\usepackage{textcomp}
\usepackage[absolute,showboxes]{textpos}

\title{Multilingual Query-by-Example Keyword Spotting with Metric Learning and Phoneme-to-Embedding Mapping}
\name{Paul M. Reuter$^{1,2}$, Christian Rollwage$^2$, Bernd T. Meyer$^1$}

\address{
  $^1$ \small{Carl von Ossietzky University, Oldenburg, Germany, 
  Communication Acoustics and Cluster of Excellence Hearing4all} \\
  $^2$ \small{Fraunhofer Institute for Digital Media Technology IDMT, Division Hearing, Speech and Audio Technology, Oldenburg, Germany}}

\email{paul.maria.reuter@idmt.fraunhofer.de, christian.rollwage@idmt.fraunhofer.de, bernd.meyer@uol.de}

\TPMargin{5pt}
\newcommand{\copyrightstatement}{
\begin{textblock}{0.84}(0.08,0.95)    
  \noindent
  \footnotesize
  \textcopyright \, 2023 IEEE. Personal use of this material is permitted. Permission from IEEE must be obtained for all other uses, in any current or future media, including reprinting/republishing this material for advertising or promotional purposes, creating new collective works, for resale or redistribution to servers or lists, or reuse of any copyrighted component of this work in other works.
\end{textblock}
}

\begin{document}

\maketitle
\copyrightstatement

\begin{abstract}
In this paper, we propose a multilingual query-by-example keyword spotting (KWS) system based on a residual neural network. The model is trained as a classifier on a multilingual keyword dataset extracted from Common Voice sentences and fine-tuned using circle loss. We demonstrate the generalization ability of the model to new languages and report a mean reduction in EER of 59.2\,\% for previously seen and 47.9\,\% for unseen languages compared to a competitive baseline. We show that the word embeddings learned by the KWS model can be accurately predicted from the phoneme sequences using a simple LSTM model. Our system achieves a promising accuracy for streaming keyword spotting and keyword search on Common Voice audio using just 5 examples per keyword. Experiments on the Hey-Snips dataset show a good performance with a false negative rate of 5.4\,\% at only 0.1\,false alarms per hour. 
\end{abstract}


\section{Introduction}

The ubiquitous presence of voice-based assistants calls for robust methods to invoke such systems using a keyword. The task of detecting the utterance of a keyword is called keyword spotting (KWS). Many approaches for keyword spotting are based on large-vocabulary continuous speech recognition (LVCSR), where traditional systems decode the speech signal and search the keyword in generated lattices \cite{miller_rapid_2007}. However, a disadvantage of LVCSR-based systems is that they often require extended computational resources and introduce a relatively high latency making them unsuitable for application on small electronic devices \cite{lopez-espejo_deep_2022}. 
This is where small-footprint systems for on-device keyword spotting come into play. One of the first keyword spotting systems and a lighter alternative to LVCSR is the keyword/filler hidden Markov model (HMM) \cite{rohlicek_continuous_1989, rose_hidden_1990}. It consists of a keyword HMM trained to model predefined keywords and a filler HMM modelling non-keyword audio segments. Viterbi decoding is used to find the best path in the decoding graph which can be computationally demanding depending on the HMM topology \cite{lopez-espejo_deep_2022}. 
In recent years, deep neural neural networks (DNNs) trained as word classifiers have shown to outperform standard HMM-based systems for KWS \cite{chen_small-footprint_2014}. Most existing works \cite{chen_small-footprint_2014, choi_temporal_2019, majumdar_matchboxnet_2020, rybakov_streaming_2020} tackle KWS as a closed-set classification problem, i.e., target and non-target keywords are predefined. Setting custom target keywords usually requires a retraining of the model with sufficient data. To address this limitation, we look for a method that allows users to set their own keywords in any language, a desirable feature for personalization of voice assistants.
A promising approach for user-defined KWS is query-by-example (QbE). In QbE, the user records a few examples of a target keyword that are compared to the incoming audio to detect the keyword. 
%
One approach for QbE is the use of phoneme posterior probabilities combined with dynamic time warping to determine the similarity between keyword samples and a test utterance  \cite{hazen_query-by-example_2009}.
The first deep neural network for Query-by-Example KWS was proposed in \cite{chen_query-by-example_2015}. The authors train an LSTM model as a word classifier and stack the outputs of the last hidden layer to obtain a fixed-length feature vector (embedding) for audio of any length. The similarity between a keyword and a test feature vector is calculated with the cosine distance. Other approaches use connectionist temporal classification (CTC) loss to train small-footprint ASR models \cite{lugosch_donut_2018, kim_query-by-example_2019}. In \cite{lugosch_donut_2018}, beam search is applied to estimate a set of label sequence hypotheses based on phonetic posteriorgrams output by the ASR model, while in \cite{kim_query-by-example_2019}, posteriograms are used to build a hypothesis graph of a finite-state transducer. 
In \cite{mazumder_few-shot_2021}, a few-shot learning method for KWS was proposed. The authors fine-tune a multilingual embedding model on a specific keyword with just five training examples and demonstrate good generalization to new languages highlighting the value of crowd-sourced data. 
Recently, metric learning has shown promising results in the user-defined KWS task \cite{huh_metric_2021, huang_query-by-example_2021, karpov_learning_2021, wang_text_2021}, aiming to directly optimize similarity in the embedding space 
without the requirement of retraining. 

There are also Query-by-String approaches where the target keyword is provided in text form~\cite{liu_rnn-t_2021}. 
A text input might be more convenient in some cases but runs the risk that the model does not match the users pronunciation of the keyword.

In this paper, we propose a system for multilingual query-by-example keyword spotting based on a residual neural network. 
The main contributions of our work are as follows:
First, we combine for the first time the use of multilingual crowd-sourced speech data and metric learning for keyword spotting, which could improve separability of the word embeddings and increase robustness and accuracy of the final system. 
Our method is compared to the classifier-based method proposed in \cite{mazumder_few-shot_2021} to demonstrate its effectiveness. 
Second, we show that the embeddings learned by the KWS model can be predicted from the phoneme sequences of the words with high accuracy using a simple LSTM architecture. 
%
Third, we explore the streaming performance of our system for keyword spotting and keyword search on Common-Voice audio \cite{ardila_common_2020} in different languages and on the publicly available Hey-Snips dataset \cite{coucke_efficient_2019}.

\section{Keyword spotting system}
\subsection{Input features and model architecture}

The inputs to the model are 40 dimensional Mel-filterbanks extracted with a Hann window of 25\,ms width and 10\,ms steps from 1\,s audio. Each input is max-normalized and log-transformed. 

Fast-ResNet-34 from \cite{chung_defence_2020} was chosen as model architecture, which is based on ResNet-34 \cite{he_deep_2016}. Residual networks allow for easier optimization of very deep architectures by introducing shortcut connections and learning residual mappings instead of original mappings. Fast-ResNet-34 uses only a quarter of the channels in each residual block hence having only 1.4\,M parameters compared to 22\,M of ResNet-34 and can be regarded as small-footprint. The full network architecture can be seen in Table~\ref{tab:fastresnet}. To produce an utterance-level fixed-length feature vector, we use temporal average pooling. 

\begin{table}[h]
	\centering
	\caption{Fast-Resnet-34 architecture. Output dimensions for an input of $ 1 \times 40 \times T $.}
	\label{tab:fastresnet}
	\scalebox{0.8}{
		\begin{tabular}{c|c|c}
		    \hline

			\textbf{Layer} & \textbf{Parameters} & \textbf{Output Size}  \\ 
			\hline

			conv1 & $7\times7$, 16, stride $2\times 1$ & $16\times20\times T$\\
			\hline
		    conv2 & $\begin{bmatrix}
                    3\times3, 16\\
                    3\times3, 16
                    \end{bmatrix} \times 3$, stride 1
                    & $16\times20\times T $\\
                    
			\hline
		    conv3 & $\begin{bmatrix}
                    3\times3, 32\\
                    3\times3, 32
                    \end{bmatrix} \times 4$, stride 2 
                    & $32\times10\times \frac{T}{2} $\\
                    
			\hline
		    conv4 & $\begin{bmatrix}
                    3\times3, 64\\
                    3\times3, 64
                    \end{bmatrix} \times 6$, stride 2 
                    & $64\times5\times \frac{T}{4} $\\
            
			\hline
		    conv5 & $\begin{bmatrix}
                    3\times3, 128\\
                    3\times3, 128
                    \end{bmatrix} \times 3$, stride 1
                    & $128\times5\times \frac{T}{4} $\\
            \hline
             & mean across frequency dimension & $128\times1\times \frac{T}{4} $ \\
            
            \hline
            
            TAP & - &  $128\times 1 $ \\
            
            \hline
            
			fc & 256 & $256\times 1$\\
			
		\end{tabular}}
		
\end{table}

\subsection{Training data}

We ran the Montreal Forced Aligner \cite{mcauliffe_montreal_2017} on the crowd-sourced Common Voice dataset \cite{ardila_common_2020} and used the alignment results to extract all words with more than three characters and a maximum length of 1 second. For training the KWS model, all words from the languages English, German, French and Catalan that had more than 500 samples were selected. The maximum number of samples per word in the dataset was set to 10,000. 25 samples per word were used for validation during training.

\begin{table}[htb!]
	\centering
	\caption{Training dataset}.
	\label{tab:cv_train}
	\scalebox{0.8}{
		\begin{tabular}{l|r|r}
		    Language &  \# words &  \# samples \\
		    \hline
		    English & 2014 & 3,694,626 \\

		    German & 703 & 1,400,878 \\

		    French & 630 & 1,130,341 \\
	    
		    Catalan & 570 & 887,858 \\
		    \hline
		    total & 3917 & 7,113,703\\
		\end{tabular}}
\end{table}

To train the model for detecting isolated keywords as well as keywords in spoken sentences, we included each extraction once padded with silence and once with its surrounding audio context in the dataset as suggested in \cite{mazumder_few-shot_2021}. By that, the number of samples was doubled to approximately 14 M. Since extractions with audio context were likely to contain more than one word, the keyword corresponding to a file label was centered in the middle of the 1\,s audio. Hence, the model learns to primarily extract information from that time region.

\subsection{Data augmentation}
\label{cha:aug}
To increase the amount and diversity of the training data we made use of various data augmentation methods. To add speaking rate variability, audio signals were resampled with a factor from 0.85 to 1.15 (as proposed in \cite{rybakov_streaming_2020}). 
Second, a random time shift in the range of $-0.05$ to $+0.05$ seconds was applied to each file. 
For additive noise and room impulse response simulation, we followed the procedure from \cite{snyder_x-vectors_2018, heo_clova_2020} for speaker verification. It was randomly chosen from one of the following categories: 
Speech, music, noise, room simulation or no augmentation. 
Noise signals were sampled from the MUSAN corpus \cite{snyder_musan_2015}, simulated RIRs originated from \cite{ko_study_2017}.

\subsection{Training details}
\label{cha:details}
Experiments on a development dataset had shown that a classification training \emph{prior} to the metric loss training improves model performance. This was implemented by adding a fully-connected layer with an output size of $N=3917$ trained for 40 epochs using cross-entropy loss with an initial learning rate of $0.001$. The learning rate was reduced after the 10th epoch using a cosine annealing schedule. We used a balanced batch sampling with batch size 128. 
After classification training, the classification layer was removed. We froze the weights of layer conv1 to conv4 and fine-tuned conv5 and fc using circle loss \cite{sun_circle_2020}, a generalization of the triplet loss function that re-weights the within- and between-class similarity scores to learn at different paces.
As suggested in \cite{sun_circle_2020}, the hyperparameters of the loss function were set to $\gamma=80$ and $m=0.4$. We adopted a P-K sampling strategy using $P=32$ (samples per class) and $K=5$ (classes). The model was trained for 10 epochs with the learning rate starting at $0.001$ and decaying after the 3rd epoch, again using a cosine annealing schedule. For both training procedures, the Adam optimizer \cite{kingma_adam_2017} was used.

\subsection{Phoneme-to-Embedding mapping} 

The requirement of recording audio examples of a target keyword might be inconvenient for some use cases. Thus, we explored the possibility of directly mapping the phoneme sequence of a target keyword to a robust word embedding learned by the KWS model. The phoneme sequence of a target keyword can be looked up in a pronunciation dictionary. A simple LSTM model was trained (as summarized in Table~\ref{tab:p2e_arc}) to predict the embedding output of the KWS model for a given phoneme sequence. The model embeds each phoneme into a 128-dimensional vector and processes the sequence with two LSTM layers. The mean of the hidden state vectors of the second LSTM is fed into a fully-connected layer to  predict the embedding. The model was trained using cosine loss. As targets, we used the mean word embeddings of 50 recordings of approx. 7,000 English words. The number of different phonemes in the training dataset was 69. 

\begin{table}[h]
	\centering
	\caption{Phoneme-to-Embedding model architecture for an input of $N$ phonemes. The embedding layer is a look-up table for 69 English phonemes and a \enquote{0} entry for padding in training.}
	\label{tab:p2e_arc}
	\scalebox{0.8}{
		\begin{tabular}{c|c|c}
		    \hline

			\textbf{Layer} & \textbf{Parameters} & \textbf{Output Size}  \\ 
			\hline

			Embedding & $70 \times 128$ & $128 \times N$\\

            \hline
            LSTM ($\times2$) & input 128, hidden 256 & $256 \times N $\\
            
             \hline
             & mean of hidden state vectors & $256 \times 1 $ \\
            
            \hline
            
			fc & 256 & $256\times 1$\\
			
		\end{tabular}}
		
\end{table}

\section{Experiments}

\subsection{Baseline}
The method proposed in \cite{mazumder_few-shot_2021} was chosen as baseline: A 3-class softmax layer (with target, unknown and background categories) is added to a pre-trained keyword classification model and fine-tuned to a specific target keyword using 5 examples. We used the pre-trained classification model from Section\,\ref{cha:details}. The target and non-target keyword samples were randomly augmented following the procedure described in Section~\ref{cha:aug}. 
Following \cite{mazumder_few-shot_2021}, we used a total of 256 training samples, with approx. 45\% augmented target samples, 45\% non-target samples and 10\% background noise from the Google Speech Commands dataset \cite{warden_speech_2018}. 
The non-target samples were drawn from a precomputed bank of 5000 keyword utterances from English, German, French and Catalan. 
In contrast to \cite{mazumder_few-shot_2021}, model weights were not frozen, which resulted in small improvements of the baseline.
The model was fine-tuned on each target keyword for 10 epochs with a batch size of 12 and an initial learning rate of $0.001$ multiplied with a factor of $0.7$ after every epoch. In the experiments, the same example and test recordings were used for the proposed method and baseline. 

\subsection{Classification accuracy}
\label{cha:class_acc}

We tested the classification accuracy of our KWS model on words not previously seen in training, which were either spoken in a language used to train the model (out-of-vocabulary words) or produced in languages unseen in training (out-of-embedding words) similar to \cite{mazumder_few-shot_2021}. The out-of-vocabulary dataset consisted of 500 words per language with 50 silence-padded extractions each. For the out-of-embedding dataset, the alignments from \cite{mazumder_few-shot_2021} were used to extract 100 words from the languages Portuguese, Turkish, Arabic and Indonesian with 25 silence-padded samples per word.
In the experiment, each word in the dataset was once used as target keyword. Five recordings of the target keyword were randomly selected as examples to the model. The remaining recordings of the word were used as positive samples and all recordings of all other words from the same language as negative samples. The embeddings of the test samples were compared to the mean embedding of the keyword examples using cosine similarity. 

In a second experiment, we varied the number of example recordings of the target keyword between one and twenty to explore the impact on classification accuracy following the same procedure. The results were compared to the accuracy achieved by predicting the target keyword embedding from the phoneme sequence using the phoneme-to-embedding model. The experiment was performed for the english out-of-vocabulary words. These were previously excluded from P2E training.

\subsection{Streaming accuracy}

In practice, keyword-spotting systems operate on a continuous stream of audio. Therefore, we investigated the streaming accuracy of our model in two contexts: Spotting isolated keywords (wake words or commands) in continuous speech (keyword spotting) and searching for keywords in spoken sentences (keyword search). Once again, classification accuracy of out-of-vocabulary and out-of-embedding words was considered and the experiments were repeated for each keyword under evaluation.
(1) We simulated wake word or command interaction by concatenating 20 randomly selected recordings of the target keyword with 200 random Common Voice sentences from the same language. Five of the remaining recordings of the target keyword were randomly selected as examples to the model. (2) To search for keywords in continuous speech, we concatenated 20 Common Voice sentences that contained the target keyword with 200 random Common Voice sentences from the same language. Five recordings of the target keyword that were not extracted from the test sentences were randomly selected as examples to the model. This procedure is similar to that in \cite{mazumder_few-shot_2021}.

We used the publicly available Hey-Snips dataset \cite{coucke_efficient_2019} 
to test the recognition performance of our model for a realistic wake word. Since the model expects the target keyword to be located roughly in the middle of an example recording, three \enquote{Hey Snips} utterances from 100 speakers (57 m, 43 f) were manually extracted.
In each run, three recordings from one speaker were chosen as examples for enrollment. For testing, we concatenated all 297 \enquote{Hey Snips} utterances from the remaining speakers and 1000 random general sentences from the Hey-Snips dataset. This procedure was repeated for every speaker resulting in a total of 121.4 hours of test audio.

\begin{figure}[htb]
    \centering
    \caption{Streaming mode} 
    \label{fig:streaming}
    \includegraphics[width=0.7\linewidth]{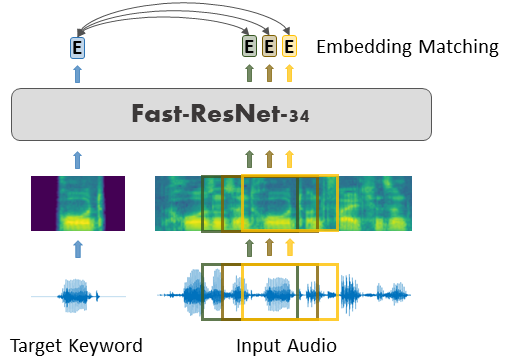}
\end{figure}

To operate on a stream of audio, a sliding window (Figure~\ref{fig:streaming}) of length 1\,s and stride 0.1\,s was used. A similarity exceeding a chosen threshold resulted in a detection. If the system output a detection, the output was suppressed for 1\,s to prevent multiple detections. Following the evaluation described in \cite{warden_speech_2018}, the the time tolerance for how close a detection must be to the ground truth’s time to count as a match was set to $0.75$\,s.

\section{Results}

In this section, we summarize our classification and streaming accuracy evaluations. Alignments with errors, sub-words or plurals of words (\enquote{refugee}, \enquote{refugees}) or phonetically similar words (\enquote{prays} and \enquote{praise}) were not manually excluded from the Common Voice test datasets and may increase the false alarm rate.

\begin{figure*}[ht]
    \centering
    \caption{(a) Classification accuracy of the proposed method and baseline for 500 out-of-vocabulary and (b) 100 out-of-embedding words per language using 5 examples per keyword. (c) Classification accuracy for 500 English out-of-vocabulary words for different numbers of examples and Phoneme-to-Embedding mapping (p2e). (d) Streaming accuracy on keyword spotting and (e) keyword search for 500 out-of-vocabulary and 100 out-of-embedding words per language using 5 examples per keyword. (f) Streaming accuracy on the Hey-Snips dataset using three examples of one speaker in every run.}
    \includegraphics[width=0.9\linewidth]{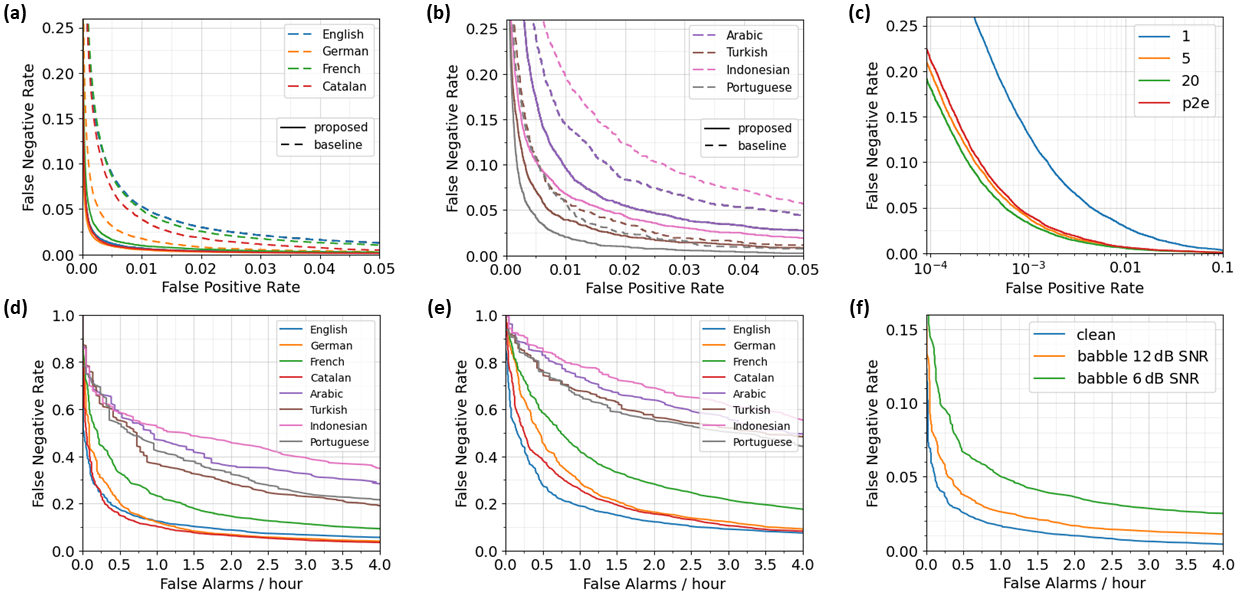} 
    \label{fig:results}
\end{figure*}

\subsection{Classification accuracy results}

We evaluate (1) the classification accuracy of the proposed method and compare it to our implementation of \cite{mazumder_few-shot_2021} and (2) the proposed Phoneme-to-Embedding mapping.

\subsubsection{Model accuracy and baseline comparison}

The classification accuracy of our model for out-of-embedding and out-of-vocabulary words is depicted in Figure~\ref{fig:results}\,a and b. We observe high accuracy for all languages seen in training~(a). Although there is a slight drop in performance, the model generalizes well to unknown languages~(b). We find that the proposed method with circle loss fine-tuning achieves higher classification accuracy than the baseline for every language in the test dataset and report a relative reduction in mean EER of 59.2\,\% from 1.96\,\% EER to 0.82\,\% for out-of-vocabulary (a) and 47.9\,\% from 3.76\,\% EER to 2.00\,\% for out-of-embedding words (b). 
The results support the findings of \cite{huh_metric_2021} that KWS systems representing non-target words as a single \enquote{unknown} class perform poorly on unseen non-target words. Instead of enforcing similarity between potentially infinite types of sounds it might be more beneficial to directly minimize similarity of non-target and target keywords with metric learning as it was done here.

%

\subsubsection{Phoneme-to-Embedding mapping}

Figure~\ref{fig:results}\,c depicts the classification accuracy of our model for different numbers of example recordings of a target keyword compared to  Phoneme-to-Embedding mapping (p2e). As expected, the accuracy improves with increasing number of examples. There is a noticeable increase in accuracy when going from one to five examples. Since further examples have less of a benefit, recording five keyword utterances seems to offer a good trade-off between user effort and accuracy. The proposed Phoneme-to-Embedding model shows to be very capable of predicting embeddings learned by the KWS model. In fact, the accuracy lies close to that for five recorded examples and P2E is a good option when less examples are available, it is however limited to the set of phonemes it was trained on. The experiment indicates that the KWS model clearly encodes phoneme-like information in the embeddings.  

\subsection{Streaming accuracy results}

We evaluate the streaming accuracy of the proposed method (1) on keyword spotting and keyword search using Common Voice  audio and (2) on the publicly available Hey-Snips dataset \cite{coucke_efficient_2019}. 

\subsubsection{Keyword spotting and keyword search on Common Voice}

Figure~\ref{fig:results}\,d reports the accuracy for spotting isolated keywords in continuous speech,  Figure~\ref{fig:results}\,e for searching keywords in spoken sentences. As expected, the former task produces better results since there are no directly interfering non-target words. For keyword search, the gap between languages seen during training vs. languages not seen becomes clearly noticeable. 

\subsubsection{Hey-Snips}

Figure~\ref{fig:results}\,f depicts the streaming accuracy of our model for the wake word \enquote{Hey Snips} for clean audio and audio with babble noise at 6 and 12 \,dB SNR. The error rates are much lower than in Figure~\ref{fig:results}\,d due the aforementioned sources of errors and the wake word being longer and less mistakable. We report a FNR of 1.7\,\% at 1\,FA per hour and an FNR of 5.4\,\% at 0.1\,FA per hour for clean audio.

\section{Conclusions}

In this paper, we proposed an approach for multilingual query-by-example keyword spotting using circle loss fine-tuning. Our model outperforms the classifier-based baseline and shows good generalization to unknown languages. We demonstrate that the learned word embeddings can be accurately predicted from the word phoneme sequences using a simple LSTM model. We achieve promising performance for streaming keyword spotting and keyword search and a competitive accuracy for a realistic wake word. The false alarms per hour reported here apply to a speech-only scenario and are expected to be much lower in an everyday acoustic environment. In future work, we will evaluate the performance of the KWS system in real acoustic environments on a target device.

\section{Acknowledgements}
This work was funded by the Deutsche Forschungsgemeinschaft (DFG, German Research Foundation) under Germany's Excellence Strategy – EXC 2177/1 - Project ID 390895286 and by - Project-ID 352015383 - SFB 1330. 
\newpage

\bibliographystyle{IEEEtran}

\bibliography{IEEEabrv,paper_kws}



\end{document}